# Advanced Asymmetrical Supercapacitors Based on Graphene Hybrid Materials

Hailiang Wang, Yongye Liang, Tissaphern Mirfakhrai, Zhuo Chen, Hernan Sanchez Casalongue, and Hongjie Dai*

*Department of Chemistry, Stanford University, Stanford, CA 94305, USA*

**Supercapacitors operating in aqueous solutions are low cost energy storage devices with high cycling stability and fast charging and discharging capabilities, but have suffered from low energy densities. Here, we grow $Ni(OH)_2$ nanoplates and $RuO_2$ nanoparticles on high quality graphene sheets to maximize the specific capacitances of these materials. We then pair up a $Ni(OH)_2$/graphene electrode with a $RuO_2$/graphene electrode to afford a high performance asymmetrical supercapacitor with high energy and power densities operating in aqueous solutions at a voltage of ~1.5V. The asymmetrical supercapacitor exhibits significantly higher energy densities than symmetrical $RuO_2$-$RuO_2$ supercapacitors and asymmetrical supercapacitors based on either $RuO_2$-carbon or $Ni(OH)_2$-carbon electrode pairs. A high energy density of ~48Wh/kg at a power density of ~0.23kW/kg, and a high power density of ~21kW/kg at an energy density of ~14Wh/kg have been achieved with our $Ni(OH)_2$/graphene and $RuO_2$/graphene asymmetrical supercapacitor. Thus, pairing up metal-oxide/graphene and metal-hydroxide/graphene hybrid materials for asymmetrical supercapacitors represents a new approach to high performance energy storage.**

Supercapacitors are recognized as an important type of devices for next generation energy storage due to their high power densities and excellent cycling stability.[1-6] They can be used as power supplies where fast and high power delivery and uptake are needed such as emergency doors on air planes. They can also provide transient power assistance to batteries and fuel cells in electric vehicles. There are two types of supercapacitors depending on energy storage mechanisms, namely electrical double layer (EDL) capacitors based on ion adsorption and pseudocapacitors based on electrochemical redox reactions.[1,5,6] The latter generally exhibits much higher specific capacitances than the former and has been used to build asymmetrical supercapacitors with improved energy and power densities.[1,6-13] Nevertheless, the energy densities of supercapacitors operating in low cost and safe aqueous solutions remain low compared to batteries.

Graphene has emerged as a promising material in energy storage and conversion applications due to its high surface area and electrical conductance.[14-24] Chemically derived graphene has shown significantly higher specific capacities than traditionally used graphite as anode materials for lithium ion batteries.[14,15] Graphene sheets obtained by reducing graphene oxide (GO) are also promising materials for double layer supercapacitors.[16-18] On the other hand, graphene is an ideal substrate for growing battery or supercapacitor electrode nanomaterials to increase capacity and rate performance by facilitating electron transfer.[19-22] Interactions with graphene could also help to control the nanocrystalline morphology of active materials and increase stability over cycling.[19-23] Recently we have synthesized hexagonal crystalline nanoplates of $Ni(OH)_2$ on graphene sheets and demonstrated high specific capacitances at high rates with this hybrid material.[22,23]

Here, we fabricated asymmetrical supercapacitors by developing a $RuO_2$/graphene hybrid material and combining it with a $Ni(OH)_2$/graphene electrode. The asymmetrical supercapacitor thus obtained showed a high energy density of ~48Wh/kg at a power density of ~0.23kW/kg, and a high power density of ~21kW/kg at an energy density of ~14Wh/kg, all with a 1.5V cell voltage operating in low-cost and safe (non-inflammable) 1M KOH aqueous solutions. The asymmetrical supercapacitor delivered higher energy and power densities than some of the best reported aqueous based supercapacitors including ones based on $RuO_2$ (one of the highest energy capacity supercapacitor materials). This is the first time that nickel hydroxide and ruthenium oxide nanomaterials are paired up to produce supercapacitors with high energy and power densities.

The $Ni(OH)_2$/graphene hybrid was synthesized by a two-step solution phase approach,[23] affording single-crystalline hexagonal $Ni(OH)_2$ nanoplates (thickness <10nm) on chemically exfoliated graphene sheets (GS, Fig. 1a) that are ideal for high capacity supercapacitor applications.[22] Note that the graphene sheets used in this paper were made by a mild chemical exfoliation-intercalation method to result in graphene sheets with significantly higher quality (lower oxygen content and higher electrical conductivity) than commonly used reduced graphene oxide.[22,23,25] The $Ni(OH)_2$/graphene hybrid material showed high specific capacitance and rate capability [~855F/g at 5mV/s and ~560F/g at 40mV/s measured by cyclic voltammetry (CV), based on the total mass of the hybrid] within a voltage range of 0-0.55V vs. a Ag/AgCl reference electrode (Fig. 1b-1d).

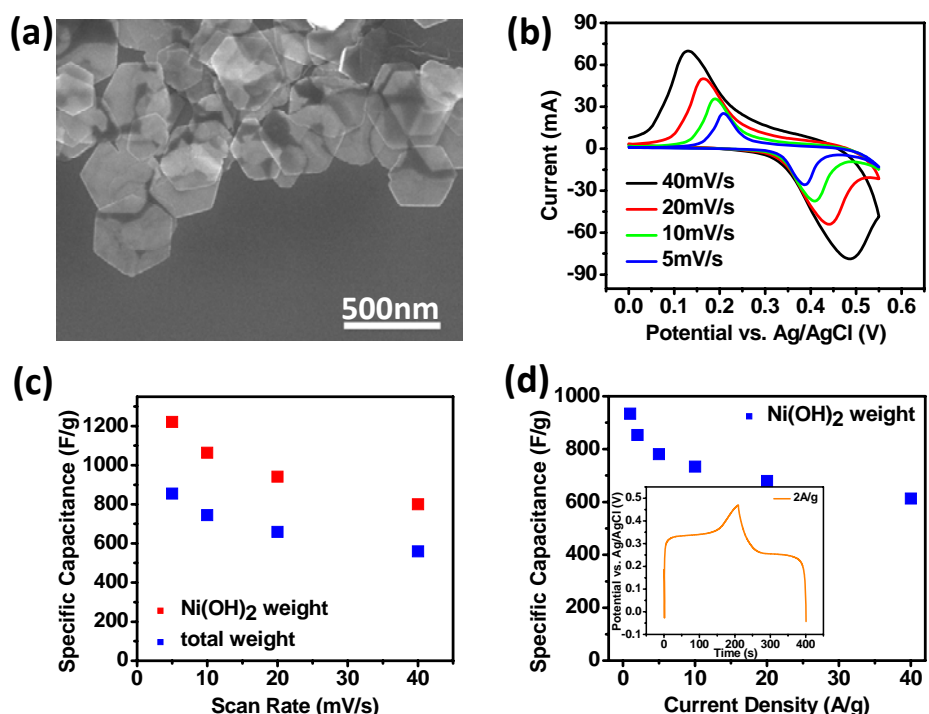

**Figure 1.** $Ni(OH)_2$/graphene hybrid material. (a) SEM image of $Ni(OH)_2$ nanoplates grown on graphene sheets. (b) CV curves of $Ni(OH)_2$/graphene hybrid at various scan rates in 1M KOH. (c) Average specific capacitance of $Ni(OH)_2$/graphene hybrid at various scan rates based on CV data. (d) Average specific capacitance of $Ni(OH)_2$/graphene hybrid at various galvanostatic charge and discharge current densities. Inset shows a galvanostatic charge and discharge curve of $Ni(OH)_2$/graphene hybrid at a current density of 2A/g.

It is highly desirable to develop a counter electrode material with similarly high performance to combine with the $Ni(OH)_2$/graphene for supercapacitors with a wider voltage range and thus higher energy density than with either material alone. Here, we developed a $RuO_2$/graphene hybrid electrode to couple with the $Ni(OH)_2$/graphene material. $RuO_2$/graphene hybrid was made in two steps (see ESM for details) in solution phase (Fig. 2a), affording small $RuO_2$ nanoparticles selectively grown on graphene sheets (Fig. 2b, 2c). The size of the $RuO_2$ nanoparticles grown on graphene was less than 10nm in diameter, revealed by the scanning electron microscopy (SEM) and transmission electron microscopy (TEM) (Fig. 2b, 2c).

The $RuO_2$/graphene hybrid material was first characterized by CV measurements against a Ag/AgCl reference electrode in a 0.2V to -1.0V voltage range in 1M KOH aqueous solutions (Fig. 2d). The $RuO_2$/graphene hybrid showed a specific capacitance of ~367F/g (based on the total mass of the hybrid) at a scan rate of 2mV/s. The specific capacitance decreased to ~280F/g when the scan rate was increased to 40mV/s (Fig. 2e, blue data points), suggesting good rate capability of the $RuO_2$/graphene hybrid material. Since the hybrid contained ~33 wt% of graphene sheets, the specific capacitances based on the mass of $RuO_2$ only would be ~1.5 X higher (Fig. 2e, red data points). The specific capacitance values of our $RuO_2$/graphene hybrid are highly competitive to other $RuO_2$ materials measured in alkaline electrolytes.[26-28] Although higher capacitances have been measured for $RuO_2$ in acidic electrolytes,[21,29-33] alkaline electrolytes are required by the $Ni(OH)_2$/graphene hybrid, which is the counter electrode of $RuO_2$/graphene in the asymmetrical supercapacitor shown below.

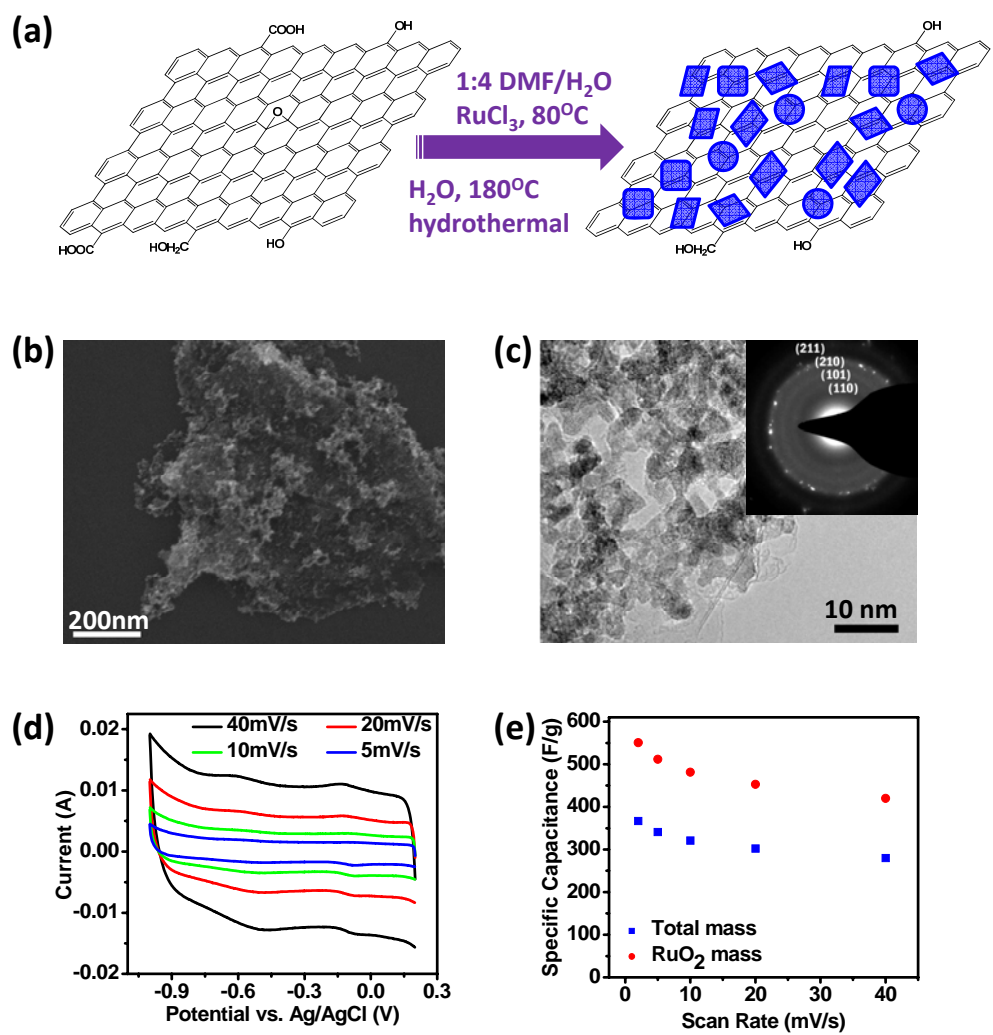

**Figure 2.** $RuO_2$/graphene hybrid material. (a) A schematic two-step growth of $RuO_2$ nanoparticles on graphene sheets. (b) SEM image of $RuO_2$ nanoparticles on graphene sheets. (c) TEM image of $RuO_2$ nanoparticles on graphene sheets. Inset shows the electron diffraction pattern of the $RuO_2$ nanoparticles on graphene. (d) CV curves of the $RuO_2$/graphene hybrid at various scan rates in 1M KOH. (e) Average specific capacitance of the $RuO_2$/graphene hybrid at various scan rates based on CV data.

Electrochemical measurements of an asymmetrical supercapacitor composed of a $Ni(OH)_2$/graphene electrode (~1mg) and a $RuO_2$/graphene electrode (~1mg) were carried out in a two electrode configuration in a beaker cell. The hybrid materials were loaded onto Ni foam substrates with 2 wt% polytetrafluoroethylene (PTFE) mixed in as binder. The mass ratio of the two electrodes were chosen to make the capacitance of the $Ni(OH)_2$/graphene electrode to be ~2 times of that of the $RuO_2$/graphene electrode, intended for a voltage drop of ~0.5V at the $Ni(OH)_2$/graphene electrode and ~1V at the $RuO_2$/graphene electrode to reach a total cell voltage of ~1.5V. Note that the specific capacitance and current density values of the supercapacitors reported below are all based on the total mass of the two electrodes excluding the Ni foam support.

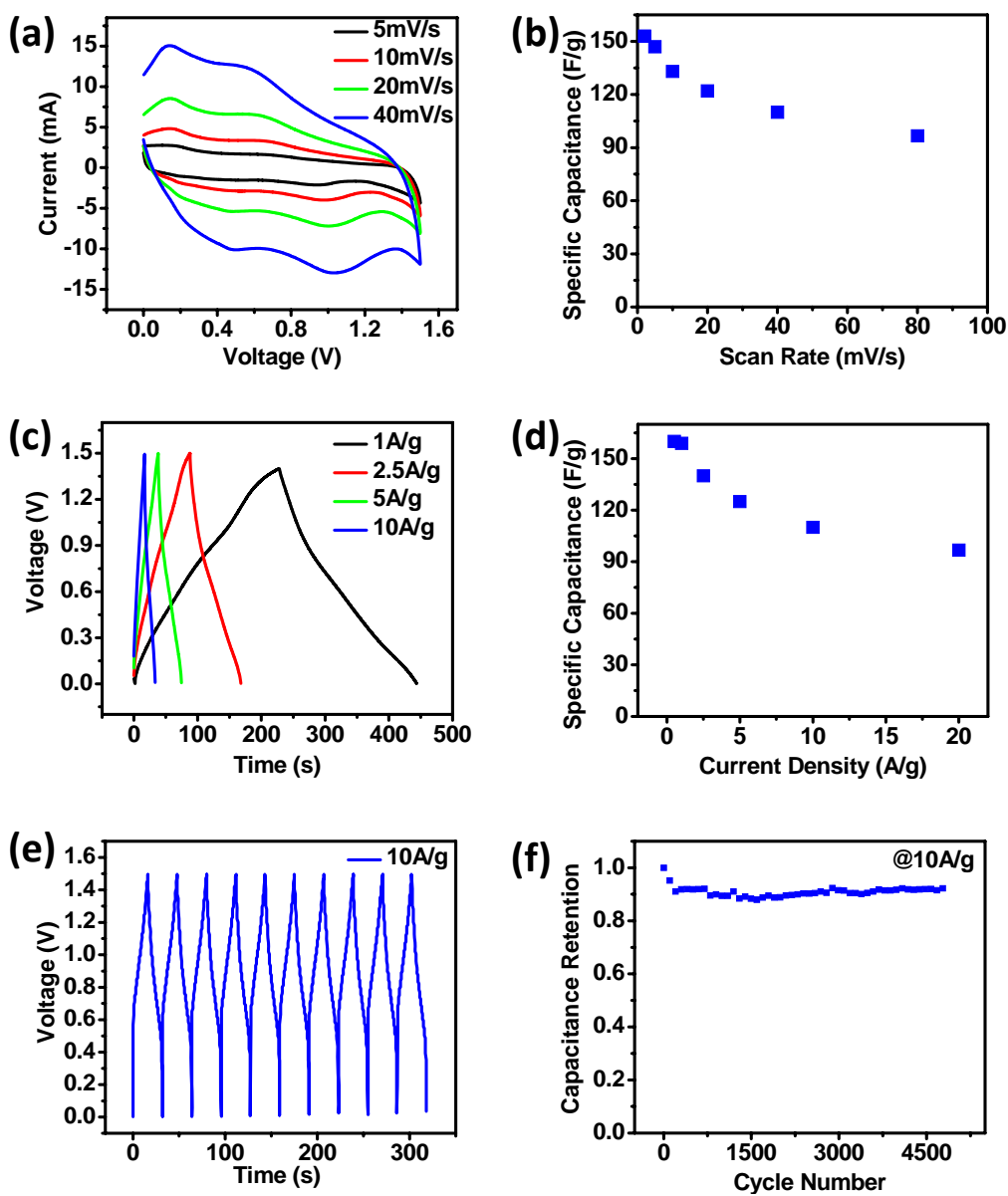

**Figure 3.** Electrochemical characterizations of the asymmetrical supercapacitor made of Ni(OH)$_2$/graphene and RuO$_2$/graphene. (a) CV curves at various scan rates in two electrode configuration in 1M KOH. (b) Two-electrode specific capacitance (based on CV data) of the asymmetrical supercapacitor at various scan rates. (c) Galvanostatic charge and discharge curves at various current densities. (d) Two-electrode specific capacitance (based on galvanostatic data) of the asymmetrical supercapacitor at various discharge current densities. (e) Galvanostatic charge and discharge curves of the asymmetrical supercapacitor at a current density of 10A/g. (f) Capacitance retention versus cycle number for the asymmetrical supercapacitor at a galvanostatic charge and discharge current density of 10A/g.

Figure 3a showed the CV curves (in a 1.5V window in two electrode configuration) of an asymmetrical supercapacitor comprised of Ni(OH)$_2$/graphene and RuO$_2$/graphene electrodes. A total specific capacitance of ~153F/g was obtained at a low scan rate of 2mV/s (Fig. 3b), based on the total mass of Ni(OH)$_2$/graphene and RuO$_2$/graphene hybrids. The capacitance was still ~97F/g at a high scan rate of 80mV/s (Fig. 3b). The high specific capacitance and rate capability of the asymmetrical supercapacitor were also revealed by galvanostatic measurements (Fig. 3c, 3d). The specific capacitance was ~160F/g at a current density of 0.5A/g. At a high current density of 20A/g, a useful capacitance of ~97F/g was measured. Interestingly, the charge-discharge voltage profiles of our asymmetrical supercapacitor were similar to those of electric double layer supercapacitors (Fig. 3c), despite of the battery-like charging and discharging behaviour of the Ni(OH)$_2$/graphene material by itself (Fig. 1b, 1d inset).[22] The asymmetrical supercapacitor also exhibited good cycling stability (Fig. 3e, 3f) with a stable capacitance (~92% of the original capacitance) after ~5000 cycles of charging and discharging at a current density of 10A/g. The Coulombic efficiency was close to 100% during cycling. Importantly, the asymmetrical supercapacitor exhibited a high energy density of ~48Wh/kg at a power density of ~0.23kW/kg, and a high power density of ~21kW/kg at an energy density of ~14Wh/kg (Fig. 4).

The high performance of our asymmetrical supercapacitor was attributed to the advanced hybrid electrode materials and their unique pairing. Both $Ni(OH)_2$ and $RuO_2$ are pseudocapacitor materials with high theoretical specific capacitance. The growth of active $Ni(OH)_2$ and $RuO_2$ nanomaterials on highly conducting graphene sheets affords optimal specific capacitance of these materials, especially in aqueous electrolytes. Pairing of the high performance $Ni(OH)_2$/graphene and $RuO_2$/graphene hybrid electrodes also expands the operating voltage of the asymmetrical supercapacitor to ~1.5V. In contrast, the maximum operating voltage range of symmetrical supercapacitors made of $RuO_2$-based electrodes is ~1V in 1M KOH aqueous electrolyte,[26,27] limited by the stability of the electrode at higher potentials. All these lead to much higher energy densities of our asymmetrical supercapacitors made of $Ni(OH)_2$/graphene and $RuO_2$/graphene hybrid materials than symmetrical capacitors made of either of the materials alone or asymmetrical supercapacitors comprised of a hybrid material electrode and a high surface area pure carbon electrode (also with a ~1.5V operating voltage), as demonstrated below.

In a control experiment, we made an asymmetrical supercapacitor by pairing ~1mg of $Ni(OH)_2$/graphene hybrid and ~3mg of reduced Hummers' graphene oxide (RGO, see ESM for preparation of the RGO electrode).[34-36] The specific capacitance of RGO was ~157F/g at a scan rate of 5mV/s (Fig. S1a, S1b), which was already good as an EDL capacitor material.[16-18] Nevertheless, it was still much lower than those of the hybrids [~1/6 of that of $Ni(OH)_2$/graphene]. The asymmetrical supercapacitor made of $Ni(OH)_2$/graphene and RGO paired electrodes (in a 1.5V voltage range) showed a specific capacitance of ~87F/g, ~74F/g and ~47F/g at a current density of 0.5A/g, 1A/g and 10A/g respectively (Fig. S2). The energy density was ~31Wh/kg at a power density of ~0.42kW/kg, and ~17Wh/kg at ~7.8kW/kg (Fig. 4). These performances were comparable to asymmetrical supercapacitors made of $Ni(OH)_2$ and activated carbon reported in literature,[7,11,13] but significantly lower than the asymmetrical supercapacitor made of $Ni(OH)_2$/graphene and $RuO_2$/graphene hybrids.

Note that a symmetrical supercapacitor made of two identical RGO electrodes showed even lower specific capacitances (only ~33F/g at 0.17A/g, Fig. S1), with energy densities of ~9.1Wh/kg and ~6.7Wh/kg at power densities of ~0.12kW/kg and ~3.3kW/kg respectively, much worse than the asymmetrical supercapacitor made of $Ni(OH)_2$/graphene paired with $RuO_2$/graphene, or $Ni(OH)_2$/graphene paired with RGO (Fig. 4). The energy density of our $Ni(OH)_2$/graphene and $RuO_2$/graphene asymmetrical supercapacitor was also significantly higher than any other pure graphene based symmetrical supercapacitors in aqueous electrolytes.[16-18]

The paring of a battery-like $Ni(OH)_2$/graphene electrode with a capacitor-like $RuO_2$/graphene electrode with non-overlapping voltage ranges afforded high capacitance to the asymmetrical supercapacitor. In a control experiment, we made a symmetrical supercapacitor using two $RuO_2$/graphene electrodes and measured a specific capacitance of ~77F/g at a current density of 0.5A/g (Fig. S3). The specific capacitance (~77F/g) was ~ 1/4 of the capacitance of a single $RuO_2$/graphene electrode (~367F/g) due to two capacitor electrodes in series with 2X of the electrode mass,[4] resulting in a low energy density of ~11Wh/kg at a power density of 0.076kW/kg (Fig. 4). Interestingly, we observed the asymmetrical supercapacitor made of $Ni(OH)_2$/graphene and $RuO_2$/graphene exhibited a specific capacitance ~1/2 (~153F/g) of the capacitance of a single $RuO_2$/graphene electrode (~367F/g). The $Ni(OH)_2$/graphene electrode was battery like, charged or discharged with sharp threshold-like voltages (see CV curves in Fig. 1b). Therefore, the voltage drops on the $Ni(OH)_2$/graphene and $RuO_2$/graphene electrode sides were not divided by simple capacitor in series model. This differs from a symmetrical $RuO_2$/graphene device in which the voltage drop is evenly divided by the two electrodes at any point of time during charge-discharge.

RuO$_2$ has been a promising high energy density supercapacitor material with a caveat of high cost.[21,26-33] Compared to RuO$_2$ based supercapacitors reported previously,[21,27-33] including RuO$_2$/graphene-RuO$_2$/graphene pair (Ref. 21) and RuO$_2$-RuO$_2$ pair (Ref. 31), our asymmetrical Ni(OH)$_2$/graphene and RuO$_2$/graphene supercapacitor shows higher electrochemical performances especially with higher energy densities (Fig. 4). Importantly, this improved performance in energy density comes with the economic benefit of replacing one of the RuO$_2$ electrodes with a low-cost Ni(OH)$_2$ electrode, greatly reducing the amount of RuO$_2$ needed for energy storage (by ~2/3 compared to RuO$_2$-RuO$_2$ devices[31]) and thus achieving higher energy densities at lower cost. It is the first time that Ni(OH)$_2$ is used to pair up with RuO$_2$ in our graphene-hybrid based supercapacitors.

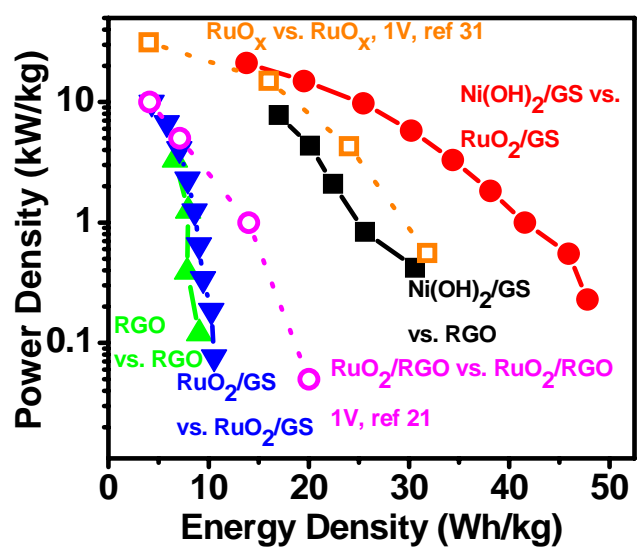

**Figure 4.** Ragone plot (power density versus energy density) of the asymmetrical supercapacitor made of Ni(OH)$_2$/graphene and RuO$_2$/graphene hybrids. Comparisons of supercapacitors composed of Ni(OH)$_2$/graphene and RGO paired electrodes (black), RuO$_2$/graphene and RuO$_2$/graphene pair (blue), RGO and RGO pair (green) together with two reference data (dashed lines).

In conclusion, we fabricated asymmetrical supercapacitors by coupling novel Ni(OH)$_2$/graphene and RuO$_2$/graphene hybrid materials. With the unique paring of battery-like Ni(OH)$_2$/graphene with capacitor-like RuO$_2$/graphene electrodes, the asymmetrical supercapacitor showed high specific capacitances and high energy and power densities with a ~1.5V operating voltage in low cost and high safety KOH aqueous electrolytes. This is the first time that nickel hydroxide and ruthenium oxide nanomaterials are paired up to produce supercapacitors with high energy and power densities. The performance of our supercapacitor reported here exceeds some of the best supercapacitors reported operating in aqueous electrolytes.